%% file: main.tex
\documentclass[conference]{IEEEtran}
\IEEEoverridecommandlockouts
\usepackage{cite}
\usepackage{comment}
\usepackage{graphicx} 
\usepackage{subcaption}
\usepackage{amsmath,amssymb,amsfonts}
\usepackage{algorithm}
\usepackage{algpseudocode}
\usepackage{graphicx}
\usepackage{textcomp}
\usepackage{hyperref}
\usepackage{xcolor}
\usepackage{booktabs}
\usepackage{soul}

\def\BibTeX{{\rm B\kern-.05em{\sc i\kern-.025em b}\kern-.08em
    T\kern-.1667em\lower.7ex\hbox{E}\kern-.125emX}}

\newtheorem{insight}{\textbf{Insight}}

\usepackage{tikz}

\begin{document}

\title{Towards Efficient Key-Value Cache Management for Prefix Prefilling in LLM Inference \thanks{This paper has been accepted at IEEE Cloud 2025. The final version will appear in IEEE Xplore.}
}

\newcommand{\ibmwatson}{IBM\,T.\,J.\,Watson Research Center, Yorktown Heights, NY, USA}

\newcommand{\emails}{yue.zhu@ibm.com, 
yuh@us.ibm.com, chen.wang1@ibm.com, zhuoran.liu@ibm.com, eunkyung.lee@us.ibm.com}

\author{
    \IEEEauthorblockN{Yue Zhu, Hao Yu, Chen Wang, Zhuoran Liu, Eun Kyung Lee}
    \IEEEauthorblockA{\ibmwatson\\
    \emails
    }
}

\maketitle

\begin{abstract}

The increasing adoption of large language models (LLMs) with extended context windows necessitates efficient Key-Value Cache (KVC) management to optimize inference performance. 
Inference workloads like Retrieval-Augmented Generation (RAG) and agents exhibit high cache reusability, making efficient caching critical to reducing redundancy and improving speed.
We analyze real-world KVC access patterns using publicly available traces and evaluate commercial key-value stores like Redis and state-of-the-art RDMA-based systems (CHIME~\cite{luo2024chime_fudan} and Sherman~\cite{sherman-sigmod22-tsinghua}) for KVC metadata management. 
Our work demonstrates the lack of tailored storage solution for KVC prefilling, underscores the need for 
an efficient distributed caching system with optimized metadata management for LLM workloads, 
and provides insights into designing improved KVC management systems for scalable, low-latency inference.

\end{abstract}


\thispagestyle{empty}
\pagestyle{plain}

\input{intro}
\input{motivation}

\input{wip_exp}

\bibliographystyle{ieeetr}
\bibliography{ref}

\end{document}

%% file: intro.tex
\section{Introduction}

Large Language Models (LLMs) have shown remarkable ability in tasks like text generation, translation, and question-answering, but their attention architecture introduces significant challenges. 
The use of key-value caches (KVC) in attention layer of transformer models, while essential for efficient token generation, requires substantial memory resources. 
As the input sequence grows, size of the KVC grows linearly with respect to the input length~\cite{kvc_generation}. 
This often limits a model's ability to handle long contexts (e.g. document summarization, conversational AI) or process multiple requests simultaneously. %
Techniques like prefix prefill improves LLM performance by caching frequently used prefixes, reducing redundant computations, decreasing time to first token (TTFT), and enhancing overall throughput~\cite{prefix_caching}. These optimizations exacerbate the memory challenge by requiring additional storage for staging pre-computed KVCs. 

Numerous research efforts have focused on addressing the challenges posed by KVC. These include quantization and pruning \cite{zhang2024more} techniques to reduce memory footprint, as well as more sophisticated approaches such as Grouped-Query Attention (GQA)~\cite{ainslie2023gqa} and Sliding Window Attention (SWA)~\cite{Longformer}. 
Additionally, researchers have explored options to offload KVC to CPU memory~\cite{xu2024piepoolingcpumemory} or sophisticated hierarchical storage solutions (e.g., Mooncake~\cite{qin2025mooncake}, LMCache~\cite{cachegen-sigcomm24}) to stage KVC across different storage tiers, allowing for more efficient management of storage resources. 
Although these works rely on traditional key-value stores or single-node data structures for metadata management, they overlook the unique access patterns of prefix prefill workloads. They do not efficiently handle KVC’s high-reusbility, dominated sequential accesses with random access patterns, causing performance degradation, increased latency, and scalability bottlenecks in LLM inference.

Recent research on KVC management has leveraged distributed system features like RDMA and disaggregated storage for traditional workloads. 
State-of-the-art key-value stores like CHIME~\cite{luo2024chime_fudan} and Sherman~\cite{sherman-sigmod22-tsinghua} focus on efficient indexing in disaggregated systems. 
However, the challenges of prefix prefill workloads in distributed environments remain largely unexplored, particularly in KVC metadata management.

In this paper, we analyze KVC block access pattern from published traces for real-world LLM-serving~\cite{qin2025mooncake}. 
Our analysis reveals fundamentally different access patterns associated with KVC-usage from traditional key-value store workloads: (1) high temporal locality for recent tokens, (2) substantial initial token reusability across requests, and (3) the need for a combination of range queries and random access operations. 
Current key-value stores, such as Redis (used in MoonCake \cite{qin2025mooncake}) and FoundationDB (used in DeepSeek 3FS \cite{deepseek3fs}), are not optimized for these unique characteristics. 
Our evaluation demonstrates that there is a need to develop a KVC management system to efficiently support the combination of range queries and random accesses while leveraging the intrinsic locality in LLM inference workloads.

%% file: motivation.tex
\section{Prior Arts \& Trace Study}
\label{sec:trace}
State-of-the-art solutions such as Mooncake~\cite{qin2025mooncake}, LMCache~\cite{cachegen-sigcomm24}, IMPRESS~\cite{chen2025impress}, and DeepSeek 3FS~\cite{deepseek3fs} 
employ different strategies for KVC staging. 
Mooncake optimizes KVC retrieval and management. 
LMCache accelerates KVC transfer via compression and dynamic merging without loss in output quality. 
IMPRESS implements an importance-aware multi-tier storage system to minimize I/O latency. DeepSeek 3FS utilizes a high-performance distributed file system with a chunk-based data store to support various AI workloads, including KVC for inference and training checkpoints. 

While these solutions aim to reduce redundant KVC computation and improve TTFT, they rely on conventional metadata management approaches. 
For instance, Mooncake leverages Redis, IMPRESS employs a node-local radix tree, and DeepSeek 3FS utilizes FoundationDB. 
None are tailored for optimized metadata management to accommodate the unique access patterns and demands of prefix prefill workloads.

To gain deeper insights into the requirements of KVC management in real-world scenario, we analyzed three prefix prefill traces (i.e., conversation, tool\&agent, synthetic trace) from Mooncake's production LLM applications~\cite{qin2025mooncake}. Due to the page limit, we present tool\&agent trace here. 
The one-hour trace is composed of a sequence of LLM-serving requests, where each request contains the arrival timestamp, input/output-length, a list of block-IDs (\textit{hash id} in the trace) for KVC blocks required in the requests.

\begin{figure}[t]
    \centering
    \subfloat[\small CDF of Block Hit Ratio]{%
        \includegraphics[width=0.45\linewidth]{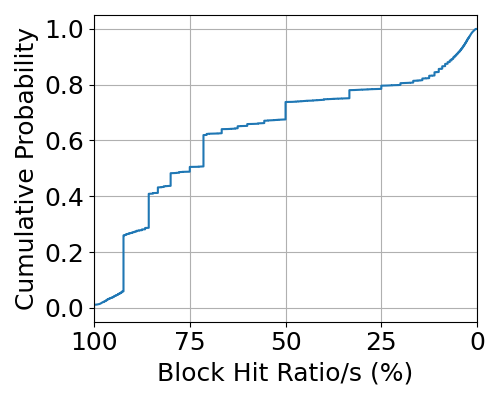}
        \label{fig:block_cdf}
        \vspace{-2pt}
    }
    \hfill
    \subfloat[\small Block ID Distribution]{%
        \includegraphics[width=0.45\linewidth]{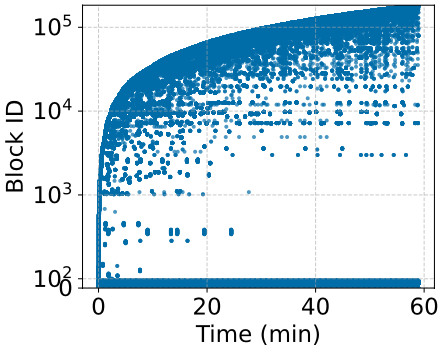}
        \label{fig:block_distr}
        \vspace{-2pt}
    }
    \caption{\small Block Reusability over 1-Hour Trace}
    \label{fig:block_hit}
    \vspace{-12pt}
\end{figure}

Using the LLM production traces, we examine the access patterns and reusability of KVC blocks. 
Fig.~\ref{fig:block_cdf} depicts the cumulative distribution of KVC-block hit rates over all KVC access requests in the trace. 
For a given request with multiple blocks, the hit-rate of the request is the ratio of the number of blocks that are accessed previously to  the total number of blocks in the request. 
As shown, over 75\% of requests have a block hit rate exceeding 50\%, demonstrating high KVC reusability. 
We saw that approximately 40\% of requests in the conversation trace and 45\% in the synthetic trace achieve a KVC block hit ratio exceeding 50\% and 80\%, respectively.
Figure~\ref{fig:block_distr} plots the distribution of KVC blocks block IDs in one hour. 
The figure shows that both the initially generated KVC and the most recently generated KVC exhibit high reusability over time. 
The same trend appears in conversation trace, 
but the synthetic trace shows rather short reuse distance in time. 

\begin{figure}[t]
    \centering
    \subfloat[\small Seq Access Rate per Req]{%
        \includegraphics[width=0.45\linewidth]{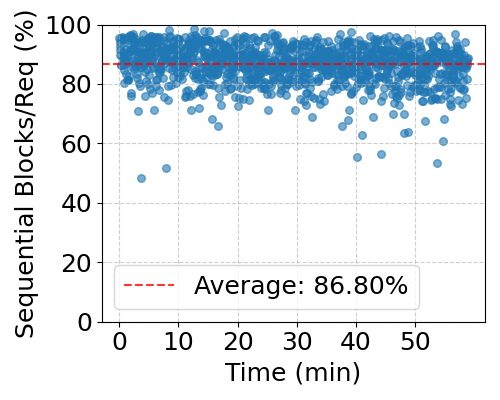}
        \label{fig:seq_access}
        \vspace{-5pt}
    }
    \hfill
    \subfloat[\small Runs Test for Non-Seq Block]{%
        \includegraphics[width=0.45\linewidth]{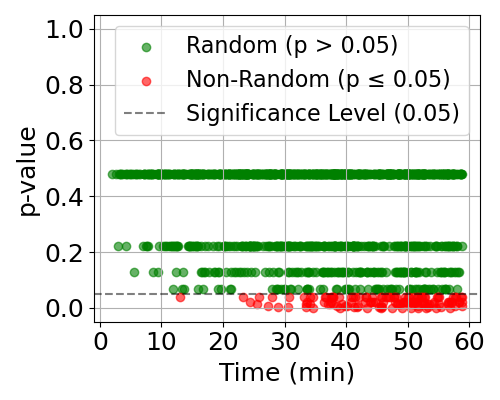}
        \label{fig:random_access}
        \vspace{-5pt}
    }
    \caption{\small Sequential \& Random Access Pattern in Requests }
    \label{fig:block_access_pattern}
    \vspace{-12pt}
\end{figure}

To further analyze block access patterns, we categorize accesses as sequential (two or more contiguous blocks) or non-sequential within each request.
Fig.~\ref{fig:seq_access} shows the fraction of sequential blocks in a request over the one-hour span. 
On average,  86.8\% of blocks within each request are sequential, enabling efficient key retrieval via range queries. By leveraging this sequential access pattern, range queries could  reduce individual lookups, minimize metadata search overhead, and improve overall retrieval latency. 
We also analyze the non-sequential block access pattern in Fig.~\ref{fig:random_access}.
We use \textit{p-values} from \textit{randomness runs test} to quantify the randomness of these non-sequential block ID distributions, where a block ID with p-value $>$ 0.05 indicates a random distribution. 
As shown in the figure, 89\% of block IDs' p-values are greater than 0.05, with many near 0.5, indicating a high random distribution of these non-sequential KVC block accesses. 
Similar sequential and random access patterns are found in the conversational and synthetic traces. 

The analysis exhibits high temporal locality for both recent and initial tokens, along with a mix of sequential and random block access. These characteristics present both opportunities and challenges for optimizing KVC management in LLM applications, highlighting the need for efficient strategies that balance sequential retrieval and handling of randomly accessed blocks.

%% file: wip_exp.tex
\section{Experiment \& Implications}

\begin{figure}[b]
    \centering
    \vspace{-13pt}
    \begin{subfigure}[b]{0.24\textwidth}
        \centering
        \includegraphics[width=\textwidth]{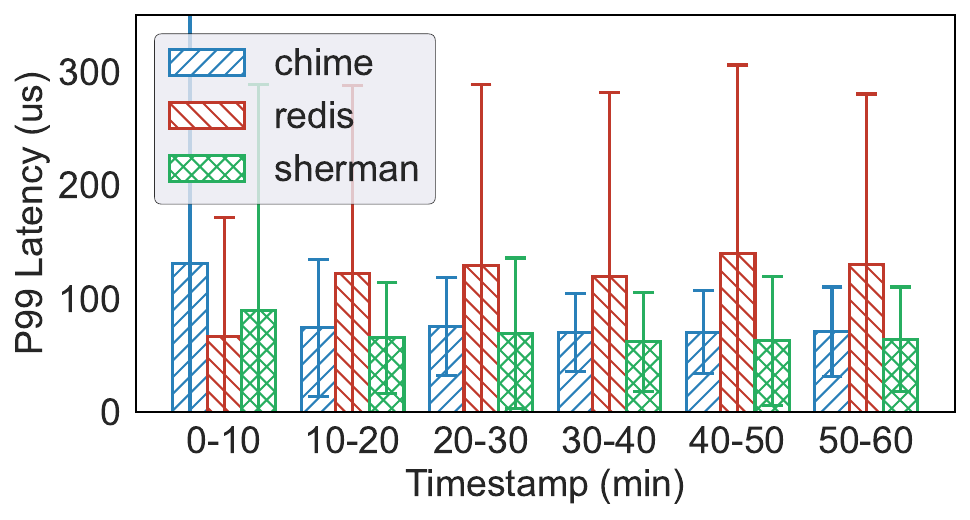}
        \caption{\small P99 Range Query Latency}
        \label{fig:range_query}
    \end{subfigure}
    \hfill
    \begin{subfigure}[b]{0.24\textwidth}
        \centering
        \includegraphics[width=\textwidth]{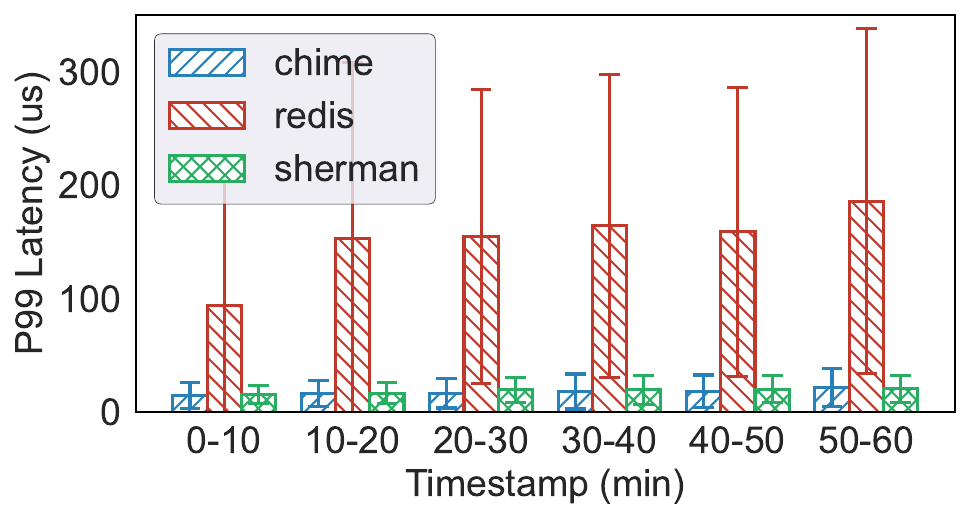}
        \caption{\small P99 Search Latency}
        \label{fig:search}
    \end{subfigure}
    \caption{\small Normalized P99 Latency Based on Real Trace (Redis=1).}
    \label{fig:exp_results}
    \vspace{-15pt}
\end{figure}

To assess the performance of commercial and state-of-the-art key-value stores in metadata management for LLM prefix prefill workloads, we conducted experiments with three systems: CHIME~\cite{luo2024chime_fudan}, Sherman~\cite{sherman-sigmod22-tsinghua}, and Redis. 
Using the publicly available KVC traces described in Section~\ref{sec:trace}, we developed a benchmark to simulate metadata operations based on the timestamp and block hash ID of each KVC block in the request stream. We assume the key is a 32-byte hash ID (SHA-256 output) and the value is an 8-byte address pointing to the corresponding KVC block. 
We issue range queries for contiguous blocks, and use \texttt{get()} operations for randomly accessed blocks. 
We conduct the experiments with two nodes, one as client and one as server, directly connected via a 100 Gbps Ethernet link. 
Each node is equipped with a 36-core Intel Xeon E5-2697 v4 CPU, 1 TiB of DRAM, and a 100 Gbps Mellanox ConnectX-6 NIC.

Fig.~\ref{fig:exp_results} presents the average p99 latency over the given time interval during the one-hour span of the real application trace,  providing insights into the performance variability of different key-value store solutions under realistic workload conditions.
In both Fig.~\ref{fig:range_query} and Fig.~\ref{fig:search}, 
Redis exhibits significantly higher latency compared to CHIME and Sherman for both range queries and search operations. 
This performance gap is primarily due to Redis's high operational overhead as a full database, whereas CHIME and Sherman are optimized for efficient metadata retrieval in distributed environments. 
Despite CHIME's reported advantages over Sherman in YCSB benchmarks, our experiments show minimal differences between the two when evaluated on KVC workloads. 
For range queries in Fig.~\ref{fig:range_query}, SHERMAN outperforms CHIME by 10.3\% on average, excluding the first 10 minutes for warm up. 
Conversely, for search latency in Fig.~\ref{fig:search}, CHIME slightly surpasses Sherman by 5.5\%, excluding the warm-up period. 
Further, both systems exhibit significant p99 latency variability in range query. 
The results suggest that the optimizations in state-of-the-art systems like CHIME and Sherman have insignificant impacts on KVC prefix prefill workloads, showcasing the need for tailored metadata management solutions that better align with the unique access patterns of these workloads. 

Based on our experiment findings and trends in KVC optimization, we outline key implications of metadata management in KVC systems tailored for prefix prefill workloads. 

\begin{insight}
{\it Traditional key-value stores were not well-suited for KVC prefix prefill workloads, which require specialized optimizations to manage both random searches and range queries while maximizing key reusability.}

When loading KVC from memory/storage stack, Redis’s long metadata indexing time (search and range query latency $>$0.1 ms) can significantly delay TTFT.
Given that the smallest TTFT ranges from 0.44 ms to 0.56 ms when pre-caching KVC on GPU memory~\cite{tangexploring}, such delays can become a major bottleneck in inference performance. 
Additionally, the structural limitations of existing systems hinder optimal KVC prefilling performance. Sherman's B+ tree is not optimized for mixed access patterns, while CHIME's hybrid approach lacks optimizations to fully exploit high key reuse. 
Their design priorities also do not align with the unique characteristics of KVC workloads. 
Sherman prioritizes write operations over read-heavy workloads, while CHIME's caching strategy fails to accommodate sequential chunk access patterns, making both suboptimal for KVC prefix prefill workloads. 
To fully leverage high key reusability while balancing random search efficiency and large-scale range queries, key-value store architectures must be reimagined, with a focus on caching mechanisms tailored to KVC access patterns. 

\end{insight}

\begin{insight}
{\it Metadata management overhead will be amplified for new KVC optimization techniques. }

Chunked prefill and KVC compression are two key optimizations aimed at improving KVC prefill efficiency. Chunked prefill (e.g., \cite{cachegen-sigcomm24}) introduces finer-granularity chunked KVC, leading to higher volume of metadata operations and increased management overhead.  
KVC compression methods, such as quantization and context-aware compression~\cite{cachegen-sigcomm24}, reduce KVC size to minimize data transfer times and enhance I/O efficiency.
As metadata operations increase and KVC data transfer times decrease, 
the relative impact of metadata operations on overall system performance becomes more pronounced. 
\end{insight}

\begin{insight}
{\it Traditional YCSB workloads is not sufficient to evaluate key-value store for KVC metadata management. }

YCSB workloads are designed for generic key-value access patterns with predefined distributions (e.g., uniform, zipfian) and fail to capture the unique characteristics of KVC metadata access, such as high key reusability and mixed random search/range query patterns. 
We will need more representative benchmarks to effectively reflect the data access patterns of KVC prefix prefill workloads. 
\end{insight}

\section{Conclusions \& Future Work}
In this work, we analyzed real-world application traces and identified the high reusability and the mixed sequential-random access patterns in KVC prefix prefill  workloads. 
To evaluate the efficiency of metadata management in existing KVC staging solutions, including Redis and state-of-the-art key-value stores, we developed a benchmark that measures metadata efficiency using real-world application traces. 
Our evaluation demonstrates that current solutions are inadequate for handling the unique demands of prefix prefill workloads. 

We are developing a metadata management system and a hierarchical KVC caching system to optimize range queries and random get queries, aiming to minimize TTFT for long-context inference via  KVC prefilling.
Our approach features a reuse-optimized metadata caching scheme, a workload-aware index structure balancing sequential block access with fast random lookups, and a hotness-aware data placement strategy for hierarchical caching.
Additionally, we will enrich the current benchmark to facilitate comprehensive evaluations of metadata management and the overall KVC caching system.